\documentclass[letterpaper]{article} 
\usepackage{aaai23}  
\usepackage{times}  
\usepackage{helvet}  
\usepackage{courier}  
\usepackage[hyphens]{url}  
\usepackage{graphicx} 
\usepackage{natbib}  
\usepackage{caption} 
\usepackage{xcolor}
\newcommand\hlbreakable[1]{\textcolor{black}{#1}}
\usepackage{algorithm}
\usepackage{algorithmic}

\usepackage{newfloat}
\usepackage{listings}
\DeclareCaptionStyle{ruled}{labelfont=normalfont,labelsep=colon,strut=off} 
\floatstyle{ruled}
\newfloat{listing}{tb}{lst}{}
\floatname{listing}{Listing}
\title{Followback Clusters, Satellite Audiences, and Bridge Nodes:\\ Coengagement Networks for the 2020 US Election}
\author{
    Andrew Beers\textsuperscript{\rm 1},
    Joseph S. Schafer\textsuperscript{\rm 1},
    Ian Kennedy\textsuperscript{\rm 2},
    Morgan Wack\textsuperscript{\rm 3},\\
    Emma S. Spiro\textsuperscript{\rm 4},
    Kate Starbird\textsuperscript{\rm 1}\\
}
\affiliations{
    \textsuperscript{\rm 1}Department of Human Centered Design and Engineering. University of Washington, Seattle, WA\\
    \textsuperscript{\rm 2}Department of Sociology. University of Washington, Seattle, WA\\
    \textsuperscript{\rm 3}Department of Political Science. University of Washington, Seattle, WA\\
    \textsuperscript{\rm 4}Information School. University of Washington, Seattle, WA\\

}

\usepackage{bibentry}

\begin{document}

\maketitle

\begin{abstract}
The 2020 United States (US) presidential election was — and has continued to be — the focus of pervasive and persistent mis- and disinformation spreading through our media ecosystems, including social media. This event has driven the collection and analysis of large, directed social network datasets, but such datasets can resist intuitive understanding. In such large datasets, the overwhelming number of nodes and edges present in typical representations create visual artifacts, such as densely overlapping edges and tightly-packed formations of low-degree nodes, which obscure many features of more practical interest. We apply a method, coengagement transformations, to convert such networks of social data into tractable images. Intuitively, this approach allows for parameterized network visualizations that make shared audiences of engaged viewers salient to viewers. Using the interpretative capabilities of this method, we perform an extensive case study of the 2020 United States presidential election on Twitter, contributing an empirical analysis of coengagement. By creating and contrasting different networks at different parameter sets, we define and characterize several structures in this discourse network, including bridging accounts, satellite audiences, and followback communities. We discuss the importance and implications of these empirical network features in this context. In addition, we release open-source code for creating coengagement networks from Twitter and other structured interaction data.
\end{abstract}

\section{Introduction}

\begin{figure}[t]
\centering
  \includegraphics[width=.45\textwidth]{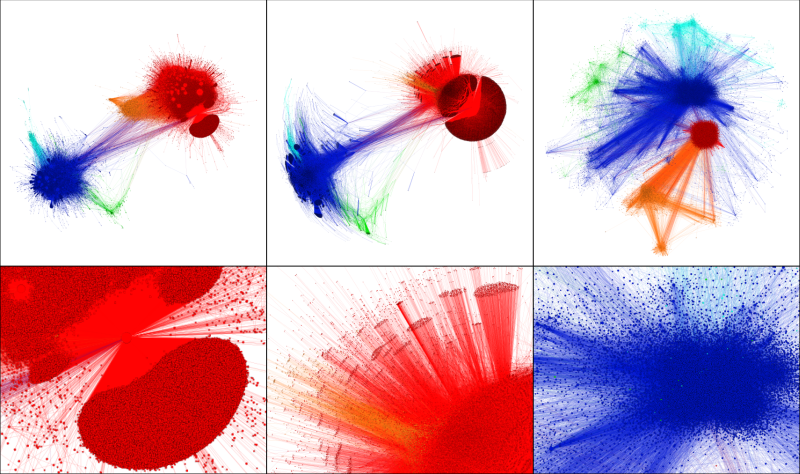}
  \caption{\hlbreakable{Directed retweet networks with typical node layout algorithms. Here, each node is an account, an edge represents a directed retweet, and edge weights represent the number of retweets. Edges with fewer than 50 retweets have been filtered out to aid visualization. The following layout algorithms as implemented in the software package Gephi are used from left to right: ForceAtlas2, YifanHu, OpenOrd. The bottom row contains close-ups of the same figures in the top row, highlighting dense node formations which obscure cluster interpretability. Dataset derived from Twitter data on the 2020 presidential election (size = 142K nodes, 424K edges).}}
  \label{fig:bad_graphs}
\end{figure}

\noindent The 2020 United States (US) presidential election was — and has continued to be — the focus of pervasive and persistent mis- and disinformation spreading through our media ecosystems, including social media \cite{center_for_an_informed_public_long_2021, benkler_mail-voter_2020}. Efforts to understand these dynamics have driven the collection and curation of large social media datasets, and the subsequent production of large, directed network representations of social interactions to make sense of them \cite{abilov_voterfraud2020_2021, kennedy_repeat_2022}. But such network datasets can resist intuitive understanding. In large network datasets, the overwhelming number of nodes and edges present in typical representations create visual artifacts, such as densely overlapping edges and tightly-packed formations of low-degree nodes, which obscure many features of more practical interest \cite{nocaj_untangling_2015, schulz_grooming_2013} (Figure \ref{fig:bad_graphs}). In the case of the US presidential election, one feature of particular interest is the functional level of interaction between different political communities who, due partly to pervasive misinformation spread in this country’s right-wing media ecosystems, no longer share a common understanding of the election’s outcome \cite{pennycook_research_2021, reuters_53_2021}. Critical to understanding these inter-community interactions is characterizing the role of platform elites, who are responsible for a disproportionate share of election-related misinformation \cite{center_for_an_informed_public_long_2021}.

Here, we present an extensive case study on a dataset of English-language Twitter posts relating to the 2020 US presidential election. We take advantage of the interpretative capabilities of coengagement networks, which are similar to the co-citation networks widely used in bibliographic scholarship. This dataset, totaling 585M retweets collected from September 1st, 2020 to December 18, 2020, contains tweets referencing generic English-language terms related to voting and the election, with a focus on tweets relating to election misinformation. In practice, this dataset contains public discourse related not only to the presidential election, but also discourse related to the persistent and false claims that the results of the election were fraudulent. We create and interrogate three different coengagement networks of retweets filtered under different parameter sets, describing via a mixed-methods analysis how the salient features of these networks correspond to different discourse phenomena. These phenomena include bridge nodes, users that are retweeted by multiple and disparate audiences; satellite audiences, groups of detached users which connect to mainstream conversations in very specific ways; and followback clusters, unique and highly active groups of users that incessantly retweet each other and very specific mainstream accounts. Our analysis of followback clusters particularly shows how Twitter’s much-noted mass account removals in the wake of the 2021 attack on the US Capitol Building particularly affected these followback groups.

Our empirical and methodological contributions together are themselves a case study in the proposed \textit{triangulation} analysis method, where the intersection of understandings from multiple, sometimes contradictory \hlbreakable{networks} generates greater knowledge than any one \hlbreakable{network} alone \cite{brandes_layout_1999, noauthor_using_1988}. We conclude this paper by discussing the advantages of coengagement \hlbreakable{networks} over other social network \hlbreakable{formats}, the importance of triangulation as a method for analysis of social networks, and future extensions and ethical considerations for using such a method.

\section{Background}
\subsection{Mis/Disinformation, Platform Elites, and US Presidential Elections}
Researchers have demonstrated the critical value of \hlbreakable{networks and network visualizations} in efforts to identify key actors in political mis- and disinformation campaigns \cite{starbird_disinformation_2019, starbird_examining_2017}. Even within work in this domain, however, the terms mis- and disinformation themselves have been variously defined \cite{jack_lexicon_nodate}, sometimes eschewed in favor of the broader term “influence operations” \cite{wanless_how_2019}, and sometimes even criticized as a contemporary moral panic \cite{mejia_white_2018, carlson_fake_2020}. For this paper, we simply define mis- and disinformation of interest as the unintentional and intentional spread of false or misleading claims that the results of the 2020 US presidential election were fraudulent. While previous research on disinformation in the 2016 US presidential election focused on foreign interference \cite{lukito_coordinating_2020}, recent analyses of disinformation in the 2020 US presidential election have focused on domestic right-wing campaigns coordinated by elites on social media and beyond \cite{center_for_an_informed_public_long_2021, benkler_mail-voter_2020}. Recent research has shown how platform elites vary between different political groupings in the US, and have highlighted their role in spreading misinformation during the COVID-19 pandemic \cite{gallagher_sustained_2021}.

\subsection{Challenges in Visualizing Social Networks}
Network visualizations of large social data can provide valuable insight into the structure of online conversations, and these visualizations have become increasingly popular as representations of computational social science’s promise \cite{foucault_welles_visualizing_2015}. A common goal in social network visualization is to highlight influential nodes and characterize the relationships they hold with one another \cite{arif_acting_2018, stewart_nobody_2021, freelon_beyond_2016}. The simplest approach is to visualize the network in its observable entirety, with nodes representing \hlbreakable{user} accounts and edges representing interactions between accounts. However, the large size of social media datasets, now often numbering in the millions or billions of nodes and edges so defined, can be intractable to visualize and render the exercise of doing so meaningless. Large social networks often encode multiple and seemingly contradictory dynamics at different scales, further exacerbating the difficulty in faithfully representing these phenomena to scholarly peers and the lay public \cite{jacomy_situating_2021}. These representational ambiguities can become especially misleading in the case of social data, where the lay-public often has strong priors about what to expect from the social world \cite{foucault_welles_visualizing_2015}. 

\hlbreakable{A common solution to this problem of large graphs is to heuristically filter unimportant nodes (e.g. with low node degree) or edges (e.g. with low edge weight) until the visualization reaches a tractable size \cite{ham_centrality_2008, dianati_unwinding_2016}. Identifying those unimportant nodes and edges is a significant challenge, as the concept of \textit{importance} is highly contingent on the interpretative aims of the researcher, and individually unimportant nodes may yet in the aggregate encode relevant structural information. Furthermore, there may be no single definition of node importance which addresses the full spectrum of phenomena represented by a social network. One innovation of particular relevance is the co-citation network developed in bibliographic network science, in which two published articles or authors are connected in a new work if other articles cite both of them together \cite{small1973co}. Importance filtering based on frequency of interaction in co-citation networks is frequently implemented, and its effects on apparent resulting clusters have been analyzed in \cite{shaw1985critical}. Co-citation networks have typically focused on scientific literature, although others have used similar principles in relation to the hyperlink structure of the web, most notably with Kleinberg's concept of online \textit{hubs} and \textit{authorities} in search engine retrieval \cite{kleinberg1999hubs}.}

\hlbreakable{Here, we extend the concept of co-citation networks to social interaction data in what we call the coengagement network. Like co-citation networks, this method significantly reduces the number of nodes visualized in large datasets, while encoding information from missing nodes in visible edges that can reveal significant relationships. Additionally, this method is also tunable, meaning that researchers can produce different visualizations according to different notions of node importance as defined by two interpretable parameters. A primary contribution is the application of co-citation principles to social data representing users, rather than documents, interacting with one another online.}

\section{Coengagement Networks}
Coengagement networks are closely related to the method of projection in bipartite graphs. In a typical bipartite projection, a network with two types of nodes and no within-group connections (such as a network composed of researchers and the papers that they author \cite{newman_scientific_2001}) is projected into one primary node type, with the remaining node type being collapsed into edge representations. This operation can productively reduce the number of nodes and edges to be visualized and analyzed, and preserves information about a primary node type while still retaining structural information from the projected node type. While we would not expect engagement networks among social media users to be bipartite, we aim to take advantage of the visual and analytical properties of bipartite projections, and therefore propose a transformation of non-bipartite graphs into bipartite graphs via the duplication of each node into two types: engaging and receiving nodes. We then project the resulting bipartite network such that engaging nodes are collapsed into edges between receiving nodes. In the context of users on Twitter, for example, this method privileges users with large audiences engaging in retweeting, commenting, liking, or even viewing, and defines relationships between users in terms of sharing similarly engaged audiences.

\begin{figure}[t]
\centering
  \includegraphics[width=.45\textwidth]{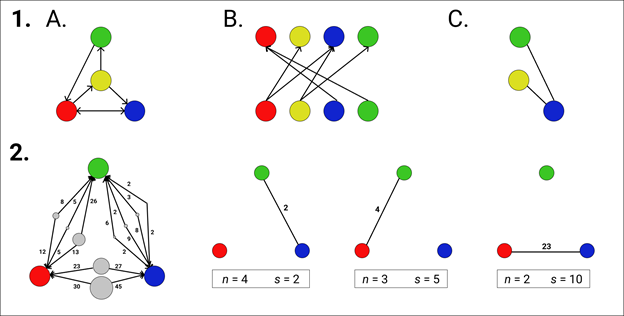}
  \caption{Schematic representation of coengagement visualizations. Panel 1) of this figure shows a schematic for transforming A) a unipartite directed graph into B) a bipartite directed graph, and then into C) a coengagement \hlbreakable{network}. The blue node is linked to the yellow node because of their shared engaging node in red, and the blue node is linked to the green node because of their shared engaging node in yellow. Panel 2) shows the effect of the node filtering parameters on an example graph. Engaging nodes, colored grey, are sized by their average out-degree. Under different combinations of the filtering parameters \textit{n} and s, different edges will result.}
  \label{fig:schematic}
\end{figure}

Formally, we define a directed graph \textit{G} = (\textit{V}, \textit{E}, \textit{w}) with vertices \textit{V}, edges \textit{E}, and edge weights \textit{w}. In the current case study, we interpret \textit{G} as a collection of Twitter users (\textit{V}) retweeting other users (\textit{E}), with edge weights w defined as the total number of retweets from one user to another. We then define a new graph \textit{G}$'$ with vertices \textit{V}$'$ , which contains duplicate sets of vertices \textit{V}$_{S}$ and \textit{V}$_{R}$ that send and receive retweets, respectively. We similarly define a new set of undirected, weighted edges \textit{E}$'$ in \textit{G}$'$, where directed edges from \textit{V}$_{i}$ to \textit{V}$_{j}$ are represented as undirected edges from \textit{V}$_{Si}$ to \textit{V}$_{Rj}$, with the same weights \textit{w}$'$. In effect, each original user has a vertex representing the instances in which they retweet others, and a separate vertex representing when they are retweeted. We then define a projection of \textit{G}$'$ as \textit{X}, such that vertices in \textit{X} are the interaction-receiving vertices \textit{V}$_{R}$, and edges in \textit{X} are defined such that the edge weight between any two vertices \textit{i}, \textit{j} in \textit{X} is the number of vertices in \textit{V}$_{S}$ that have defined edges to both \textit{V}$_{Ri}$ and \textit{V}$_{Rj}$. This final vertex set in the projected graph represents users that are retweeted and draws edges between them when they are jointly retweeted by at least one other user. We finally define two edge filtering parameters \textit{n} and \textit{s} on the resulting graph \textit{X}. Specifically, an edge between two users in \textit{X} is defined if at least \textit{n} other users have retweeted both users at least \textit{s} times each. The parameter \textit{n} represents a minimum diversity of users retweeting two users, while \textit{s} represents the minimum volume of retweeting a user must do to be considered in n.

The \textit{n} and \textit{s} parameters, which in practice control the number and distribution of edges in a coengagement \hlbreakable{network}, are a powerful tool for targeting specific visualizations. These edge filtering parameters allow researchers to shape the output of their \hlbreakable{networks} along two important, yet distinct, qualitative dimensions by modifying the distribution of edges between users. When researchers filter with a higher \textit{n} value, influential users are related only if they attract engagement from large, diverse audiences, a typical goal in influential user analysis. When researchers filter with a higher \textit{s} value, nodes are instead related by audiences that frequently retweet their content with a dataset, which can reveal dedicated rather than transient audiences. Different ratios of \textit{n} to \textit{s} can reveal other relationships: low \textit{n} with high \textit{s} can make highly active and coordinated audiences more salient, whereas low \textit{s} with high \textit{n} make more salient those infrequent instances in which content is shared widely across different communities.

In what follows, we present a series of case studies using a dataset of tweets related to the 2020 US presidential election. Specifically, we use a dataset of 585M Twitter posts containing \hlbreakable{substrings related to voting and the election (‘vote’, ‘voting’, ‘mail’, ‘ballot’, ‘poll’, and ‘election.')}, define nodes as the authors of these posts, and (unprojected) edges as retweets from one user to another. \hlbreakable{We do not include quote tweets.} We show that depending on the choice of the parameters \textit{n} and s, different clusters of influential nodes can be distinguished, and different forms of qualitative analysis can be applied. In doing so, we demonstrate the practical implications of coengagment \hlbreakable{networks} and their interpretation. Empirically, these case studies offer a unique look at engagement during the 2020 presidential election.

In the first case study, we choose a very high value of \textit{n} and \textit{s} = 1 to create a \hlbreakable{network} that shows a broadly two-part structure to Twitter discussions around the presidential election, aligned with pro-Trump and pro-Biden accounts. The low \textit{s} parameter highlights transient instances of high-volume crossover between these two groups but does not necessarily represent \textit{sustained} engagement across these groups. We then choose a parameter set with much lower \textit{n} and slightly higher s, to illustrate how a third pro-socialist grouping becomes salient at different audience sizes, and how some crossover nodes do not hold sustained engagement. We end with a third case study at very high \textit{s} and very low n, to highlight two new followback communities that become salient when active, sustained engagement is prioritized over large audiences.

\section{Findings}
\subsection{Case 1: Bridging Between Clusters at High Audience Sizes}

\begin{figure}[h]
\centering
  \includegraphics[width=.45\textwidth]{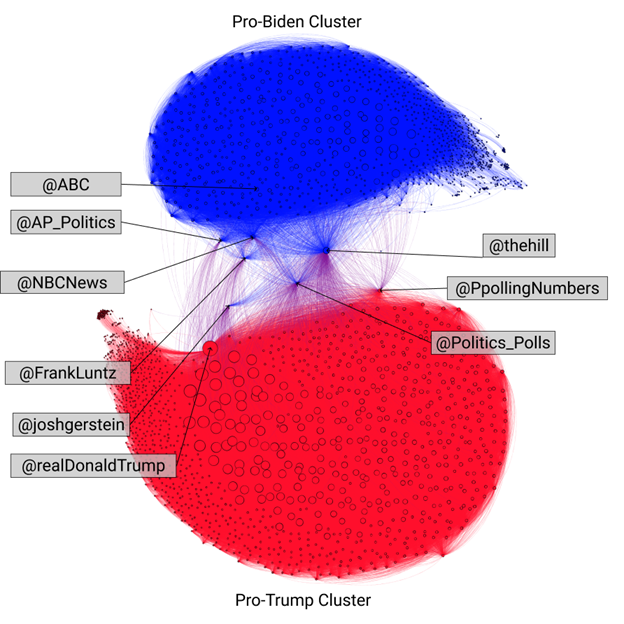}
  \caption{Case 1, election discourse for large audiences (s=1, \textit{n} = 10,000). A coengagement visualization of retweet relationships in a collection of tweets related to the US presidential election. Each node represents a Twitter account, and two nodes are linked if at least 10,000 users retweeted them both nodes at least once. Edges are undirected and weighted according to how many users retweeted both nodes. Nodes are sized according to their weighted degree, i.e. the sum of the weights of their incoming edges. Highlighted nodes represent nodes with connections to both pro-Trump and pro-Biden clusters.}
  \label{fig:case_1}
\end{figure}

\begin{figure}[t]
\centering
  \includegraphics[width=1\columnwidth]{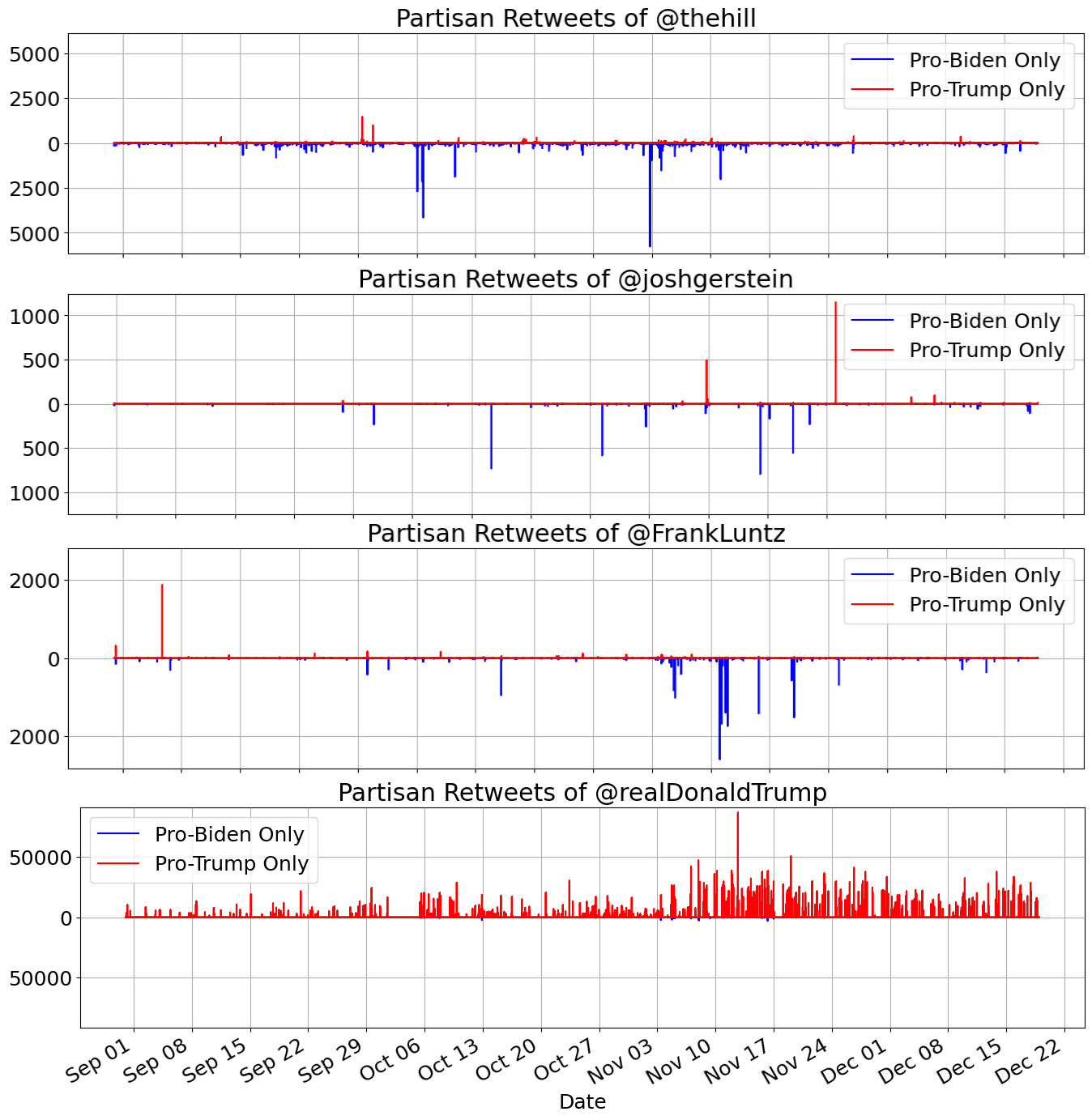}
  \caption{Fig. 4. Example accounts with bipartisan engagement in Case 1. Plots of partisan retweets of four different bridging accounts labeled in Figure 3. Tweet dates are marked by the first time they were retweeted in our dataset. “Pro-Trump Only” partisan retweets indicate users who only tweeted accounts labeled in the pro-Trump cluster in Figure 3. “Pro-Biden Only” signifies the same totals for the pro-Biden cluster. Retweets of bridging nodes themselves are excluded in both retweet totals.}
  \label{fig:case_1_line_graphs}
\end{figure}

In Case 1, we generate a coengagement \hlbreakable{network} where node relationships are defined by low restrictions on retweet frequency (\textit{s} = 1), but high restrictions on total retweet volume (\textit{n} = 10,000, Figure \ref{fig:case_1}). \hlbreakable{In this and future visualizations, we use the ForceAtlas2 \cite{jacomy_forceatlas2_2014} visualization algorithm as implemented in the application Gephi \cite{bastian_gephi_2009}.} Intuitively, this means that two nodes are connected if at least 10,000 users retweeted both at least once during the period of this dataset, and nodes with more users retweeting both of them are more tightly linked together. Visualizing this projection reveals two tightly interconnected communities of influential users discussing the US presidential election, which we term and briefly describe as pro-Trump, and pro-Biden clusters. The pro-Trump cluster is anchored around Donald Trump’s account, and includes an array of pro-Trump political activists, political organizations, politicians, media outlets, journalists, anonymous and self-identified online influencers, activists from the antifeminist “manosphere” online culture, and conspiracy-based QAnon communities. The pro-Biden cluster includes an array of politicians, journalists, media outlets, and online influencers, some of which self-identify as pro-Biden or liberal, and others of which, such as the television network CNN, identify as non-partisan but rejected false pro-Trump claims of election fraud. 

Overall, 2,499 nodes are generated in this graph, with 1,385 nodes in the pro-Trump cluster (55\%) and 1,114 nodes in the pro-Biden cluster (45\%). However, this set of retweets may be biased towards pro-Trump accounts due to our focus on terms related to election misinformation disproportionately spread by pro-Trump accounts.

We use the “pro-” descriptor to describe clusters as a whole, but some individual members of these clusters may identify themselves otherwise. We also use the terms pro-Trump and pro-Biden, rather than Republican and Democrat, to illustrate the extent to which traditional US party alignments are contradicted in the membership of these clusters. The Lincoln Project, a Republican advocacy group that supports conservative causes, is one of the most prominent accounts in the pro-Biden cluster, while many accounts in the pro-Trump cluster have ambivalent stances towards the Republican party outside of Trump. Their membership in a community is contingent not on their core beliefs or associations, but on their behavior in the dataset we observe, namely their posting behavior in the months before and after the presidential election. In datasets with the same users but different topics of discussion, the membership of individual users may change.

Audiences rarely retweet across clusters in large numbers, and when they do, they tend to retweet a select few cluster-spanning nodes. Almost all (98\%) cross-cluster connections route through only eight nodes, labelled in Figure \ref{fig:case_1}. These bridging nodes serve different functions in this discourse environment, and connect people in different ways (Figure \ref{fig:case_1_line_graphs}). The most common form of bridging node was one where the account generally created two types of tweets, one that appealed to pro-Trump audiences, and one that appealed to pro-Biden audiences. The most transparent examples of these accounts were those that tracked polling results (@PpollingResults, @Politics\_Polls, @AP\_Politics, @NBCNews, @APPolitics), where those results and polls that favored Biden were retweeted by pro-Biden accounts, and those that favored Trump were retweeted by pro-Trump accounts. However, this form of apparent bridging also occurred when media accounts reported in neutral tones on events that fed preexisting pro-Trump and pro-Biden narratives respectively. The journalist account @joshgerstein was separately retweeted by both clusters for neutrally reporting on Trump’s attempts to contest the election results, with apparent pro-Trump legal judgments being more retweeted by pro-Trump users and their subsequent legal refutations more retweeted by pro-Biden users. The alternating quality of these nodes complicates the notion that they bridge communities, as their tweets are most often disproportionately shown to only one community at a time.

A special form of this alternating bridging occurred with the account @FrankLuntz, which posted updates on polls and predictions about the outcome of the presidential election. Before election day, this account was sometimes critical of Biden and released some predictions favorable to Trump, which led it to garner a slightly right-leaning cumulative audience. After the election, this account was resolute in affirming Biden’s victory in the face of false pro-Trump claims of election fraud, then earning a growing pro-Biden audience. This account displays the sensitivity of such analyses to dataset selection, as a pre-election discourse analysis would likely place the account firmly in the pro-Trump cluster, whereas a post-election discourse in the pro-Biden, and when combined, firmly between. It was rare when accounts created posts that consistently appealed to audiences in both clusters. Some individual posts had equal appeal across clusters, such as when polling accounts released vote tallies tied at nearly 50\% in critical states. Other posts that had a similar appeal were simply neutral statements of fact about recent news relating to the presidential election, which were made particularly often by the account for the online news organization The Hill (@thehill). While pro-Biden leaning, the Hill’s account was one of the only accounts to consistently find engagement from both pro-Biden and pro-Trump accounts across this election time period.

The last significant point of crossover between the pro-Trump and pro-Biden clusters is the account for former president Trump itself (@realDonaldTrump). This circumstance reveals that though we imply for much of this analysis that retweets constitute endorsements, they do not always behave as such. Many pro-Biden accounts may be retweeting Trump’s account simply because his tweets are often consequential in and of themselves, even when they are clearly opposed to Biden’s election to the presidency. In several instances, pro-Biden users likely retweeted Trump’s tweets sarcastically, such as when pro-Biden users disproportionately retweeted an old tweet from 2012 reading: “Scary thought--@JoeBiden is a heartbeat away from the Presidency.” 

\subsection{Case 2: Third Party and Satellite Audiences at Lower Audience Sizes}

\begin{figure}[h]
\centering
  \includegraphics[width=.45\textwidth]{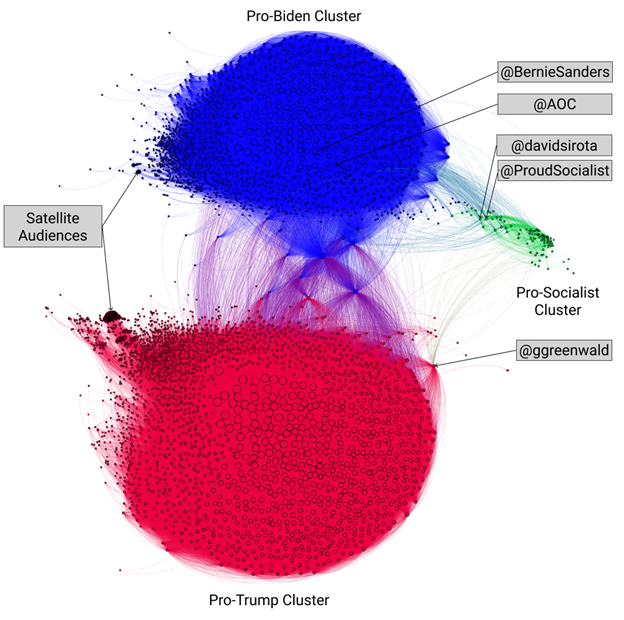}
  \caption{Case 2, Election discourse for medium audience sizes (\textit{s} = 5, \textit{n} = 100). A coengagement visualization of retweet relationships in a collection of tweets related to the US presidential election. Each node represents a Twitter account, and two nodes are linked if at least 100 users retweeted them both nodes at least five times. Edges are undirected and weighted according to how many users retweeted both nodes. Nodes are sized according to their weighted degree, i.e. the sum of the weights of their incoming edges. Nodes bridging between the pro-socialist and other clusters are labelled, as well as what we term satellite audiences: clusters of nodes with low-degree representing audiences with only tangential connection to mainstream US election discourse.}
  \label{fig:case_2}
\end{figure}

\begin{table*}
\resizebox{\textwidth}{!}{
\begin{tabular}{llllll}
\hline
\multicolumn{6}{|l|}{Top Election Accounts by Cross-Cluster   Connections (Case 2)} \\ \hline
\multicolumn{2}{|l|}{Biden-Trump} &
  \multicolumn{2}{l|}{Biden-socialist} &
  \multicolumn{2}{l|}{Trump-socialist} \\ \hline
\multicolumn{1}{|l|}{Account Name} &
  \multicolumn{1}{l|}{\begin{tabular}[c]{@{}l@{}}Share of \\Cross-Cluster\\ Connections\\     (\textit{n} = 2,515)\end{tabular}} &
  \multicolumn{1}{l|}{Account Name} &
  \multicolumn{1}{l|}{\begin{tabular}[c]{@{}l@{}}Share of \\Cross-Cluster\\ Connections\\     (\textit{n} = 299)\end{tabular}} &
  \multicolumn{1}{l|}{Account Name} &
  \multicolumn{1}{l|}{\begin{tabular}[c]{@{}l@{}}Share of \\Cross-Cluster\\ Connections\\     (\textit{n} = 34)\end{tabular}} \\ \hline
\multicolumn{1}{|l|}{thehill} &
  \multicolumn{1}{l|}{24\%} &
  \multicolumn{1}{l|}{Proudsocialist} &
  \multicolumn{1}{l|}{16\%} &
  \multicolumn{1}{l|}{ggreenwald} &
  \multicolumn{1}{l|}{79\%} \\ \hline
\multicolumn{1}{|l|}{PpollingNumbers} &
  \multicolumn{1}{l|}{18\%} &
  \multicolumn{1}{l|}{BernieSanders} &
  \multicolumn{1}{l|}{13\%} &
  \multicolumn{1}{l|}{jimmy\_dore} &
  \multicolumn{1}{l|}{9\%} \\ \hline
\multicolumn{1}{|l|}{Politics\_Polls} &
  \multicolumn{1}{l|}{11\%} &
  \multicolumn{1}{l|}{davidsirota} &
  \multicolumn{1}{l|}{12\%} &
  \multicolumn{1}{l|}{TulsiGabbard} &
  \multicolumn{1}{l|}{6\%} \\ \hline
\multicolumn{1}{|l|}{realDonaldTrump} &
  \multicolumn{1}{l|}{7\%} &
  \multicolumn{1}{l|}{AOC} &
  \multicolumn{1}{l|}{11\%} &
  \multicolumn{1}{l|}{aaronjmate} &
  \multicolumn{1}{l|}{6\%} \\ \hline
\multicolumn{1}{|l|}{AP\_Politics} &
  \multicolumn{1}{l|}{6\%} &
  \multicolumn{1}{l|}{briebriejoy} &
  \multicolumn{1}{l|}{10\%} &
  \multicolumn{1}{l|}{} &
  \multicolumn{1}{l|}{} \\ \hline
\multicolumn{1}{|l|}{threadreaderapp} &
  \multicolumn{1}{l|}{5\%} &
  \multicolumn{1}{l|}{KyleKulinski} &
  \multicolumn{1}{l|}{7\%} &
  \multicolumn{1}{l|}{} &
  \multicolumn{1}{l|}{} \\ \hline
\multicolumn{1}{|l|}{Garrett\_Archer} &
  \multicolumn{1}{l|}{5\%} &
  \multicolumn{1}{l|}{IlhanMN} &
  \multicolumn{1}{l|}{6\%} &
  \multicolumn{1}{l|}{} &
  \multicolumn{1}{l|}{} \\ \hline
\multicolumn{1}{|l|}{spectatorindex} &
  \multicolumn{1}{l|}{4\%} &
  \multicolumn{1}{l|}{ryangrim} &
  \multicolumn{1}{l|}{5\%} &
  \multicolumn{1}{l|}{} &
  \multicolumn{1}{l|}{} \\ \hline
\multicolumn{1}{|l|}{FrankLuntz} &
  \multicolumn{1}{l|}{3\%} &
  \multicolumn{1}{l|}{peterdaou} &
  \multicolumn{1}{l|}{5\%} &
  \multicolumn{1}{l|}{} &
  \multicolumn{1}{l|}{} \\ \hline
\multicolumn{1}{|l|}{DecisionDeskHQ} &
  \multicolumn{1}{l|}{3\%} &
  \multicolumn{1}{l|}{RBReich} &
  \multicolumn{1}{l|}{3\%} &
  \multicolumn{1}{l|}{} &
   \\ \hline
\end{tabular}}
\caption{Bridging connections between pro-Biden, pro-Trump and pro-socialist clusters. The top 10 Twitter accounts in terms of total cross-cluster edges for each two-cluster pairing in Case 2. These accounts share audiences of at least 100 users retweeting at least 5 times each across election discourse clusters.}
\label{table:users}
\end{table*}

In Case 1, a high \textit{n} parameter demonstrated the nodes and graph structure of accounts with a relatively high volume of retweets over this dataset. While revealing the relatively few shared points of reference between pro-Biden and pro-Trump audiences, groups of users with smaller retweet bases are not visible in this graph. To visualize these communities, we generate the same data at a dataset with lower n, instead filtering by \textit{s} to keep the size of the node and edge set tractable. Specifically, we generate a \hlbreakable{network} in which links are defined when two nodes are shared by 100 users four times (\textit{n} = 100, \textit{s} = 5, Figure \ref{fig:case_2}). After removing non-US clusters, this reveals a third group of nodes that we term the pro-socialist cluster. The pro-socialist cluster, much smaller than either the pro-Biden or pro-Trump clusters, is centered around multiple political activists, journalists, and influencers associated with US democratic socialist candidates and causes. We note that, like the pro-Biden and pro-Trump cluster, that this cluster does not contain all pro-socialist accounts, and some of its members may not identify as such. Indeed, popular democratic socialist presidential candidate Bernie Sanders is located in the pro-Biden cluster rather than pro-socialist cluster, likely due to his widespread popularity and continued public support for Biden during this election.

At a lower volume of retweeting, more bridges appear between the pro-Biden and pro-Trump clusters, and new bridges are generated between the pro-socialist and other clusters. Notably, the relative importance of some bridges changes under the new requirement for repeated engagement (\textit{s} = 5). For example, the account for reporter @joshgerstein has no cross-cutting connections in this graph, despite being responsible for 7\% of all connections in the first graph. This discrepancy is likely caused by how much of this account’s pro-Trump retweet engagement comes from only two tweets, both, describing in a neutral tone, updates on pro-Trump attempts to legally invalidate the results of the election. This change demonstrates the effect of the \textit{s} parameter, which can be tuned upward to select nodes for sustained engagement over a dataset, rather than widespread but momentary engagement in critical posts. The opposite effect can also be seen in those new nodes that appear as significant bridges. The spam media account @spectatorindex becomes a bridging node in Case 2, due to its sustained but low level of engagement, as do circumstantially important accounts like the governor of Arizona’s (@dougducey), whose state was the center of many false election fraud claims.

Bridges between the pro-socialist and other clusters illustrate the different modes of inter-cluster commonality that can exist between different user clusters in this graph (Table 1). Between the pro-Biden and pro-socialist clusters, there are many nodes that draw consistent engagement, including popular democratic socialist politicians Bernie Sanders, Alexandria Ocasio-Cortez, and Ilhan Omar, and an array of writers, podcast hosts and other online influencers associated with the socialist movement. Between the pro-Trump and pro-socialist clusters, however, there is little engagement, and all of it is mediated through three nodes. Most prominently among these nodes is Glenn Greenwald, a journalist who combined left-leaning views on topics such as surveillance with frequent engagement with right-wing media and criticism of mainstream media.

This second case study introduces a visualization feature which we had previously aimed to eliminate: tightly-packed clusters of low-degree nodes, in this case one-degree nodes mostly connected only to @realDonaldTrump. We note that while in ordinary graphs such nodes are usually uninteresting, in coengagement visualizations the relative isolation of these nodes reveals an important function in the election discourse environment. Particularly, many of these one-degree nodes are from communities plausibly isolated from mainstream US political discourse, but still displaying, for example, a contextual support for Trump or Biden. We term edges stemming from these nodes as \textit{satellite audiences}, with inspiration from \cite{squires_rethinking_2002}. 

These nodes include accounts from other countries and/or in other languages, such as high-follower right-wing accounts writing for Japanese or Brazilian audiences. Such accounts are unlikely to interact with the majority of English-language right-wing accounts prominent in this graph, but may retweet Trump as a signal of nominal allegiance to his movement. Other low-degree nodes may originate in popular English-language, US-based communities that focus on topics usually unrelated to electoral politics. For example, the low-degree account for pop musician Ariana Grande, one of the most followed accounts on Twitter, connects only to Biden and fellow musician Lady Gaga, signaling a possible separation between entertainment-focused audiences and mainstream election-focused audiences.

\subsection{Case 3: Followback Clusters at High-Frequency Engagement}

\begin{figure}[t]
\centering
  \includegraphics[width=.9\columnwidth]{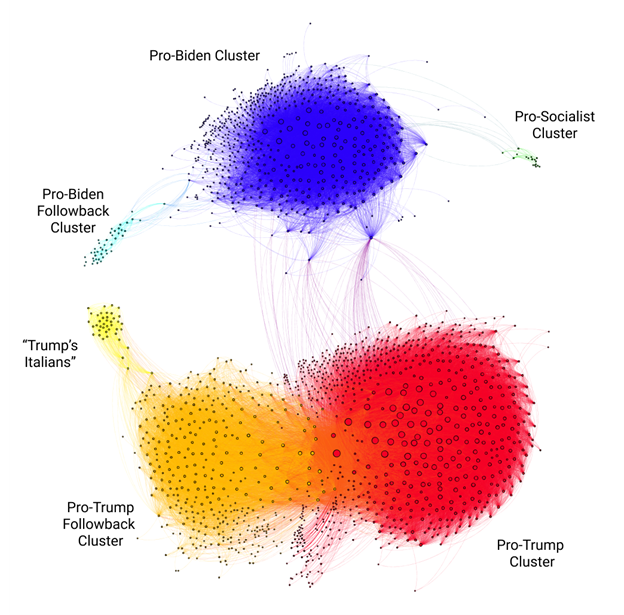}
  \caption{Case 3, Election discourse for small but active audiences (\textit{s} = 25, \textit{n} = 25). A coengagement visualization of retweet relationships in a collection of tweets related to the US presidential election. Each node represents a Twitter account, and two nodes are linked if at least 25 users retweeted them both nodes at least 25 times. Edges are undirected and weighted according to how many users retweeted both nodes. Nodes are sized according to their weighted degree, i.e. the sum of the weights of their incoming edges. Followback clusters, comprising nodes which retweet and follow other accounts to a relatively extreme degree, are labeled (Pro-Biden Followback Cluster, Pro-Trump Followback Cluster, “Trump’s Italians”).}
  \label{fig:case_3}
\end{figure}

Implicitly in the previous cases, structure is mostly determined by the number of users choosing to retweet two different accounts. However, as \textit{s} increases and \textit{n} decreases, structure is increasingly determined by repeated interactions by relatively small groups of users, which makes the actions of well-coordinated groups more salient. To illustrate this, we generate a graph where links are defined by 25 users retweeting two nodes at least 25 times each over the course of the dataset (\textit{n} = 25, \textit{s} = 25, Figure \ref{fig:case_3}). 

\begin{figure*}[t]
\centering
  \includegraphics[width=.9\textwidth]{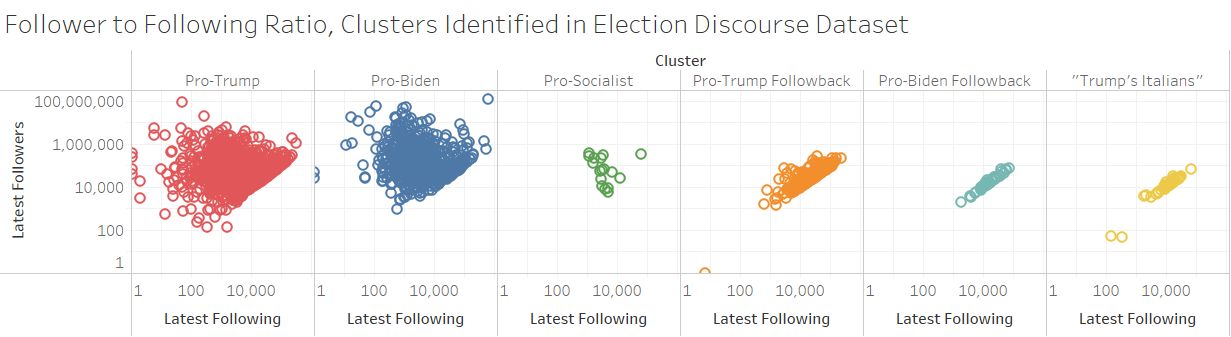}
  \caption{Follower to following ratios for Case 3 clusters. Scatter plots of the number of followers and the number of accounts followed for each Twitter account visualized in Figure 6 (n = 25, s = 25). Follower and following totals are counted at the time of the latest tweet recorded in the dataset. Plots are separated and colored according to cluster membership, and both axes are on logarithmic scale.}
  \label{fig:case_3_ratios}
\end{figure*}

In this case, three new clusters emerge with ties to the existing pro-Trump and pro-Biden clusters. We term the new clusters in this graph \textit{followback} communities, due to their unique method of gaining followers and using Twitter. Accounts in these communities attempt to gain followers by mass-following other accounts in expectation of reciprocal follows, and sometimes explicitly coordinate with other accounts to expose themselves to a wider audience of potential followers. Because Twitter limits the number of users an account can follow by that account's current follower number, these accounts can often be distinguished from others by their nearly 1:1 ratio between followers and following totals (Figure \ref{fig:case_3_ratios}). In addition to this follower manipulation practice, these communities have other unique behaviors compared to the clusters previously identified. Their median retweet total is much higher than that of the previously-identified clusters, and retweets are a much higher percentage of their total tweeting behavior. Their frequent retweeting likely propels their visibility in this visualization. Qualitatively, their behavior is also different from other users on the platform. They engage in retweet “trains,’ in which they make posts tagging members of their own community and then retweet these posts incessantly in an attempt to garner more followers for all participants \cite{gallagher_trump_2020}. They are almost entirely pseudonymous, with screen names and profile information often detached from any offline presence. One of the followback clusters is much larger than the others and associated with the core pro-Trump cluster, one is smaller and an off-shoot of the larger pro-Trump followback cluster, and the smallest cluster is associated with the core pro-Biden cluster.

As with previous clusters, we can investigate the nodes which bridge one cluster to another. In this case, however, both followback clusters have no connections to non-followback clusters not aligned with their preferred candidate. In all three cases, an important point of cross-cluster connection are the accounts for the two presidential candidates themselves (@JoeBiden and @realDonaldTrump), but unlike other groups, most nodes in the followback clusters are connected to these nodes. This feature of dense cross-cluster connectedness reflects a critical function of the engagement from followback clusters: to retweet the followback community, but also to retweet the influential nodes supporting their presidential candidate of choice. 

Multiple followback clusters can exist supporting the same presidential candidate, and different clusters may share unique user characteristics. By inspecting the usernames and user-entered profile descriptions of the smaller pro-Trump followback cluster, we found that 29/41 accounts provided some indicator of Italian American identity, either explicitly identifying as such, identifying as part of “Trump’s Italian Army,” or including Italian flag emojis paired with US flag emojis. While the larger pro-Trump followback also contained some self-identified Italian users, they were by no means the majority as in the smaller cluster. This smaller cluster was linked to the main pro-Trump followback cluster only through four bridging nodes, two of which identified as part of Trump’s Italian army, and two which self-described as duplicate accounts for the same user. This cluster otherwise only formed internal connections and connections with @realDonaldTrump and @Llinwood, the account of L. Lin Wood, a lawyer who advanced many false conspiracy theories relating to the election and litigated on Trump’s behalf in some post-election court cases.

All three followback clusters reflect the curious quality of coengagement \hlbreakable{networks} where users can play both the role of edge and node. Recall that coengagement \hlbreakable{networks} are graph projections, where each edge is comprised of engagements from many unseen nodes present in a ordinary, unprojected version of the graph. Almost 38\% of the unprojected nodes that comprise the edges in followback clusters are themselves represented as nodes in the coengagement \hlbreakable{networks}, compared to less than 1\% of the accounts in the core pro-Trump, pro-Biden, and pro-socialist clusters. This is to say that followback clusters uniquely play the role of their own audience, relentlessly sharing their own members' content in addition to their chosen candidates. This feature sheds light on how the smaller pro-Trump followback cluster can separate itself in this graph, as its self-identified Italian American accounts specifically and at the threshold we set retweet only each other. This analysis does not allow us to speculate as to the level of explicit coordination or the motives behind these accounts for acting in this way. While some of these clusters’ unusual behavior could indicate automated or semi-automated activities, previous reporting has indicated that at least some of these accounts are likely operated and coordinated by otherwise ordinary users that simply aim to support Trump’s candidacy on Twitter \cite{gallagher_trump_2020}.

\subsection{Synthesis: Cluster Contingency and Differential Moderation}

\begin{figure}[t]
\centering
  \includegraphics[width=1\columnwidth]{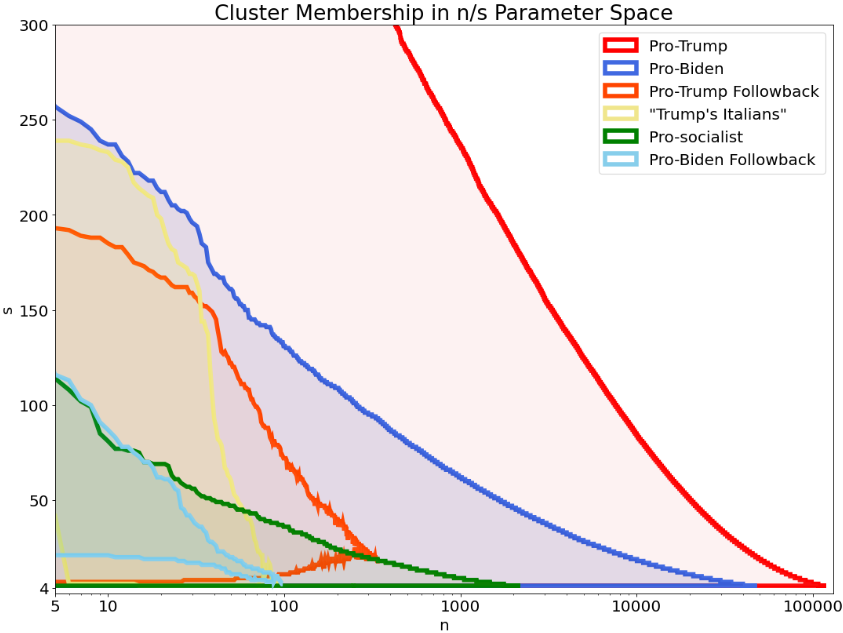}
  \caption{Existence map for clusters across filtering values n, s. For each cluster identified in these three case studies, we determine the maximum parameter values at which these clusters can be identified. Clusters are labeled at each (n, s) parameter value if they contain specific high-degree landmark nodes identified in the previous case studies. In the shaded regions, individual clusters are salient to the clustering algorithm. Outside of the shaded regions, these clusters cannot be identified either because their constituent nodes are not present or because they have been subsumed into other clusters.}
  \label{fig:meta_graph}
\end{figure}

By examining the full \textit{n/s} parameter space, we can develop a sense for the relative size and sharing characteristics of the clusters that we have identified so far. We calculate the results of repeated network clusterings using the Louvain algorithm on a range of possible \textit{n} and \textit{s} parameter combinations, assigning cluster labels via high-degree landmark nodes associated with each previously-identified cluster. In Figure \ref{fig:meta_graph}, we illustrate the boundaries at which these clusters are no longer salient. The pro-Trump, pro-Biden, and pro-socialist clusters are all contained within a maximum number of engaged users that steadily declines with restrictions on these users’ number of retweets. By contrast, the followback clusters cannot be detected unless users’ minimum retweets are elevated, but are never salient at a size of greater than 110 users. We note that the point at which a cluster fails to be detectable is not the point at which all nodes in these clusters are removed, but rather the point at which these nodes become subsumed into larger clusters. For example, the two highest degree pro-socialist nodes from Case 2 are relatively-marginal nodes in Case 1, while many of the most popular Trump followback accounts in Case 3 are found in the core pro-Trump cluster in Cases 1 and 2. These facts reiterate that the clusters we identify here are not found with respect to the actual ideologies or social ties between nodes, but rather through an understanding of how nodes are perceived by their engaged audience variously defined.

\begin{figure*}[t]
\centering
  \includegraphics[width=.9\textwidth]{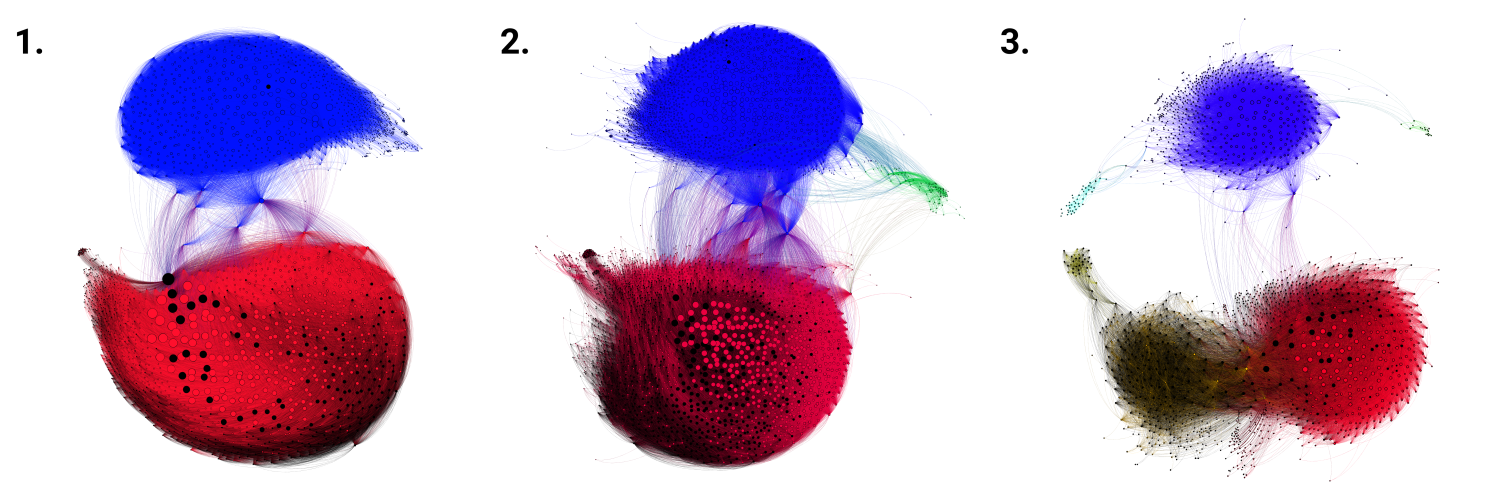}
  \caption{Visualization of suspensions across all case studies, labeled (1-3). The original networks from all three case studies are visualized in increasing order from left-to-right. Nodes that have been suspended by Twitter as of January 6, 2021 are colored in black, otherwise coloring remain the same as in the original case studies.}
  \label{fig:suspensions}
\end{figure*}

We conclude these case studies by noting how the clusters we identify here were subject to different levels of moderation in the wake of the US presidential election (Figure \ref{fig:suspensions}). On January 6, 2021, a large group of rioters, associated with a range of pro-Trump movements contesting the results of the 2020 presidential election, entered the US Capitol building while its representatives were in session. Following this event, Twitter suspended a large number of accounts said to have encouraged this violent protest, as well as accounts associated with the QAnon conspiracy movement \cite{romm_twitter_2021}. To measure the effect of these and other more recent suspensions on our dataset, we identified all accounts that were suspended as of September 3, 2021. Since the election, most of the pro-Trump followback cluster in this case study (71\%) has been suspended by Twitter, compared to 32\% of accounts in the core pro-Trump clusters, 2\% of the core pro-Biden cluster, 7\% of the pro-Biden followback cluster, and 2\% of the pro-socialist cluster. In other words, these suspensions have disproportionately affected the pro-Trump followback users we identified in Case 3, which is to say groups of pro-Trump users with frequent intracommunity retweet activity.

\subsection{Comparison: Directed Engagement Graph and Flows}

We conclude our case studies with a brief comparison to a more standard network form, which we refer to as the directed engagement graph. In directed graphs, each node is an account, each edge is a directed retweeting relationship, and edge weights signify the frequency of engagement. In this present dataset, the unfiltered directed graph consists of 22M nodes and 299M edges. This graph cannot be visualized effectively with our current software and computing power, although a version with heavily filtered edge weights can be observed in Figure 1.

\hlbreakable{We first assess whether coengagement graphs represent nodes typically perceived to be important in the directed graph. We compare the weighted (in)degrees of both graph, which in the case of the directed engagement graph is simply equal to the number of times an account has been retweeted. We find that of the top 1000 most retweeted accounts, 95\%, 96\% and 85\% are represented across Cases 1, 2, and 3 respectively. Furthermore, we find that the nodes we choose to include in Cases 1, 2, and 3 account for 54\%, 64\%, and 51\% of \textit{all} retweets in this dataset, despite only representing fewer than 0.1\% of the accounts in our data. There are some highly-retweeted nodes in the directed graph that are not visible in each coengagement network. However, these missing nodes are also mostly unrelated to the 2020 US presidential election. For example, the most retweeted nodes missing from Cases 1, 2, and 3 respectively are a fan account for pop musician Justin Beiber, a fan account for the Korean pop music group BTS, and the account for Billboard, a music media outlet. These nodes are highly retweeted in these datasets because of competitions in which fans “vote” for their favorite musicians. They may be missing from our coengagement visualization simply because they are essentially apart from the apparent main topic of the dataset, and thus have fewer opportunities to “share” an audience with another node that is collected under these terms. Some relevant highly-retweeted accounts, such as the Twitter account for Donald Trump's daughter Ivanka Trump, are also missing from Case 3 likely because they did not publish enough election-related tweets to reach the 25-retweet threshold.}

\hlbreakable{We second assess whether structures found in the directed graph are significantly different from the coengagement graph. To do this, we cluster the directed graph using the Infomap algorithm \cite{rosvall2009map}, a different clustering method which views directed, weighted edge interactions as flows of information between nodes. We find that the pro-Biden, pro-Trump, socialist, and pro-Trump followback clusters are all still found under this algorithm, with the pro-Biden followback cluster being subsumed into the pro-Biden cluster. No other large cluster is found which combines nodes found in these case studies into new mixtures. We do find, however, that clusters apparent in these case studies, such as the pro-Biden cluster, often are subdivided into smaller clusters in the directed network.}

These smaller clusters reveal the tradeoffs between directed networks and coengagement networks, as they most likely stem from Infomap's tendency to privilege connections between high-indegree nodes, i.e. retweets between influential accounts. For example, a pro-Biden subcluster in the directed graph is centered around official accounts and reporters for the media outlet The New York Times. These popular accounts frequently retweet each other, probably to promote each others' work, which strengthens their association in graph forms and algorithms which understand high-degree nodes as routes through which users travel. By contrast, in coengagement graphs, the influence of interactions between influential nodes on the resulting form and clusters is dampened, due to its focus on \textit{large} but not necessarily \textit{important} shared audiences. In social engagement data in which user engagements can be seen as travelling from node to node, such as use clicks through profile pages, directed retweet networks and clustering algorithms that consider them may be more appropriate. In data such as this Twitter retweet dataset, where engagements are enacted from relatively stable positions and circumstances, coengagement graphs may be more appropriate.

\section{Discussion}
We introduce coengagement \hlbreakable{networks} and illustrate their value for a mixed methods analysis of a dataset of Twitter posts related to the 2020 US presidential election. We illustrate how the number of apparent clusters perceived in these \hlbreakable{networks} is contingent on the minimum proposed size and activity level of their engaged audiences. When seen through the lens of large, momentarily interested audiences, there appears to be two dominant pro-Trump and pro-Biden communities. When including smaller engaged audiences, a pro-socialist cluster emerges, and when focused on highly-active but even smaller engaged audiences, unique followback clusters emerge with severely different user behavior. These \hlbreakable{networks} make clear that Twitter’s moderation in the wake of the attack on the US Capitol Building disproportionately affected these pro-Trump followback accounts, while also affecting a number of pro-Trump accounts with more ordinary retweeting behavior. Taken together, the insights from these \hlbreakable{networks} depict an ecosystem of popular and activist discourse communities in the presidential election with few but crucial points of overlap, and the \textit{de facto} removal of the majority of influential member is one of these communities in the wake of the US Capitol attacks.

The purpose of our case study is not necessarily to draw definitive conclusions or cause-and-effect relationships, but to use visualization to expose those features of this discourse ecosystem that deserve further study. We describe a mostly binary structure of pro-Trump and pro-Biden engagement in English-language US election discourse, and provide a typification of the points of crossover between these two clusters of accounts. We identify a detectable and yet marginal tide of third-party US political discourse in the pro-socialist cluster whose growth and comparative influence on non-election discourses may yet have further importance. We identify the phenomenon of satellite audiences, where high-degree nodes, and particularly the account @realDonaldTrump, serve as singular points of reference for many communities with only marginal connection to English-language election discourse. And we characterize followback communities, which use unique posting strategies and engage in unusually partisan rhetoric to support specific candidates in our election dataset.

The social network visualization approach described here makes transparent many features which contribute to this understanding of Twitter discourse around this election, while reducing visual clutter and artifacts typical to equivalent large datasets. However, we stress that the sum total narrative could not be found from any of these visualizations taken alone, and indeed some visualizations have contradictory features which could, when viewed in isolation, generate misleading insights as to the nature of these communities. For example, we have shown that the pro-Trump followback cluster is from the perspective of the size of their engaged audience a somewhat marginal phenomenon, ceasing to be organizationally coherent at the level of more than a hundred users. Portraying them with equal prominence to more typical political communities may mislead viewers as to their relative impact on the overall election conversation compared to, for example, the core pro-Trump cluster. Yet small groups of coordinated users can nonetheless have an outsize impact in online communities \cite{center_for_countering_digital_hate_disinformation_nodate}, and our analysis of Twitter suspensions since the presidential election show that this followback community was specifically targeted for moderation at a much greater rate than other communities. Depending on the visualization goals of the researcher — proportional impact, or behavioral diversity — forefronting this community may or may not be informative to their chosen audiences. Portraying multiple visualizations that provide both interpretations, and analyzing the discrepancies between them, provides a level of explanatory power that no single visualization is likely to otherwise achieve.

\hlbreakable{Overall, what we describe in this paper is an interpretative, mixed-methods workflow, in which visual artifacts derived from quantitative network transformations are combined with a deep qualitative understanding of the US Twitter election context, to the benefit of both. Given the contingency of network visualizations on their initial parameters, deep understanding of the context of a given dataset is necessary to interpret and triangulate their different incarnations. However, given the size of contemporary social interaction datasets, quantitative methods such as the coengagement network are necessary to reduce the complexity of online conversations to artifacts which are tractable to qualitative researchers. We advocate for the continued use of coengagement networks in mixed-methods research, as a tool that both stimulates and benefits from deep contextual understanding of increasingly unwieldy social data.}

\vspace{-5pt}

\subsection{Ethical Considerations, Limitations, and Software Sharing}
As network visualizations continue to be central in social media analysis, we believe it is necessary to briefly examine the ethical considerations on whether network visualizations such as these \textit{should} be used in every circumstance. We have chosen here to visualize users participating in a high-prominence topic, consisting of mostly public-facing accounts such as politicians and media outlets. The same methods applied to communities with a higher expectation of privacy, or who face higher risks from exposure, may be unethical surveillance if researchers have not derived consent from members of these communities. There are also ethical implications to naming accounts visualized as nodes in networks. Some users, due to gender, race, or other factors, are at higher risk of harassment if identified as influential in a given community, while other users who explicitly seek attention in online communities may use their identification in networks as a propaganda tool in hateful campaigns. In our reporting of this work, we have declined to name some accounts for both reasons. We also stress here how the data collection procedure, and subsequent description of that procedure, affects which communities appear to be participating in a phenomenon. Several politically-active Twitter communities in the US that have been previously described in research, such as Black Twitter \cite{clark_tweet_nodate} and non-English language communities \cite{fang_social_2021, soto-vasquez_covid-19_2020}, are not explicitly visible in our analysis, likely due to their different posting volumes and the choice of terms and topics on which we chose to center our data collection. Researchers working with such visualizations whose research bears on policy and public perception must explain such limitations in the communication of their work.

We have described the basic form of \hlbreakable{an approach} for visualizing engagements in social network data, and there are many ways in which this method can be modified to more saliently capture engagement dynamics. For example, the current formulation of this method places emphasis on users who frequently share content, which is not necessarily undesirable given the external impact of this behavior. However, different formulations of the network projection scheme, such as those that weight users’ engagements relative to their average level of engagement, may be a better reflection of real discourse communities that exist at lower sharing volumes. We also observe that many of these coengagement \hlbreakable{networks} create densely-connected subgraphs in which most nodes are connected to most other nodes, making internode relationships difficult to visually identify. Accordingly, these \hlbreakable{networks} may be complementary with other techniques to improve visualizations of social networks, such as the edge sparsification procedures for densely-connected networks proposed by Nocaj et al. (\citeyear{nocaj_untangling_2015}).

In the hopes that others may replicate our methods of both visualization and analysis on new datasets, we make the code available for generating these graphs either from structured data received from the Twitter API, or in general JSON and CSV-based formats.\footnote{\url{https://github.com/uwcip-research/Coengagement-Networks}} This code uses the visualization capabilities of the open-source network visualization library Gephi, and its implementation of the ForceAtlas2 algorithm for network visualization \cite{bastian_gephi_2009, jacomy_forceatlas2_2014}. We have packaged this code in publicly-available Docker containers, a relatively portable and stable code format which can be run on many machines with relatively few installation requirements. Additionally, we have made available node and link data for all visualizations displayed in this paper, as well as a list of Twitter ID numbers for tweets and users corresponding to data used to generate these graphs. We hope by making the code for generating these graphs open-source, other researchers both qualitative and quantitative will both explore the potential and limitations of this method, as well as contribute modifications to this scheme as appropriate.

\section{Acknowledgments and Funding}
We would like to acknowledge Clément Levallois for his advice on the open-source implementation of this project, and Paul Lockaby for ample support in data collection. We received funding from the University of Washington's The Center for an Informed Public, The John S and James L Knight Foundation (G-2019-58788), Craig Newmark Philanthropies, and the Omidyar Network.

\bibliography{CameraReady/LaTeX/aaai23}

\end{document}